\documentclass[amsmath, amssymb,12pt,letterpaper]{elsarticle}
\pdfoutput=1

\usepackage[dvips,margin=1.25in]{geometry}
\usepackage{color}
\usepackage{graphicx}
\usepackage{caption} 
\usepackage{subcaption} 
\usepackage{bm}
\usepackage{amssymb}
\usepackage{amsmath}
\usepackage{cancel}

\newcommand{\pref}[1]{(\ref{#1})}

\def\({\left(}
\def\){\right)}
\def\[{\left[}
\def\]{\right]}

\def\barray{\begin{array}}
\def\earray{\end{array}}
\def\be{\begin{equation}}
\def\ee{\end{equation}}
\def\ben{\begin{equation} \nonumber}
\def\een{\end{equation}}
\def\bea{\begin{eqnarray}}
\def\eea{\end{eqnarray}}
\def\eal{\end{align}}
\def\bal{\begin{align}}

\def\({\left(}
\def\){\right)}

\def\One{{\hbox{ 1\kern-.8mm l}}}

\newcommand{\bS}{{\mathbb S}}

\def\lstr{\ell_{\rm str}}

\begin{document}

\date{}                                           

\title{%
Fractionated Branes and Black Hole Interiors
}
\author{Emil J. Martinec}
\address{
Enrico Fermi Inst. and Dept. of Physics, University of Chicago\\
5640 S. Ellis Ave., Chicago, IL 60637-1433, USA\\
e-martinec@uchicago.edu
}


\begin{abstract}
Combining a variety of results in string theory and general relativity, a picture of the black hole interior is developed wherein spacetime caps off at an inner horizon, and the inter-horizon region is occupied by a Hagedorn gas of a very low tension state of fractionated branes.  This picture leads to natural resolutions of a variety of puzzles concerning quantum black holes.
Gravity Research Foundation 2015 Fourth Prize Award for Essays on Gravitation.
\end{abstract}

\maketitle


%




\section{Black Hole Horizons and Thermodynamics}
\label{sec:Thermo}





The outer horizon of a stationary black hole is the surface beyond which outward-directed light sheets have a negative expansion.  Causal evolution draws any matter within this surface further into the interior.  For charged and/or rotating black holes, such outward-directed light sheets accumulate at an inner horizon.
For instance, the metric of a charged black hole can be written in Eddington-Finkelstein coordinates
\be
\label{EFgeom}
ds^2 = -\Delta(r)dv^2 + 2dv \,dr + r^2 d\Omega^2  ~;
\ee
the two horizons are located at $\Delta(r_\pm)=0$.  

Properties of both horizons feed into black hole thermodynamic relations such as the first law%
~\cite{Cvetic:1997uw,Castro:2012av}:
\be
dM = T_\pm dS_\pm + \Omega^\pm dJ + \Phi_e^\pm dQ_e ~.
\ee
Here the gravitational entropy associated to the horizon is its area in Planck units; the temperature is related to the surface gravity
\be
S_\pm = \frac{A_\pm}{4G}
~~,\quad
T_\pm = \frac{\kappa_\pm}{2\pi} ~~;
\ee
$\Omega_\pm$ is the angular velocity of the horizon, and $\Phi_e^\pm$ its electrostatic potential.

Because the inner horizon is an accumulation point of outgoing lightsheets, perturbations of initial data are arbitrarily blueshifted and a shockwave develops along $r=r_-$; the inner horizon is unstable%
~\cite{Poisson:1990eh,Hiscock1981110,Marolf:2011dj}.   
One may thus call into question the analytic extension of the vacuum geometry beyond $r=r_-$ and expect that a fundamental theory regularizes the null singularity that arises there.  Indeed, in many solutions carrying the quantum numbers of extremal charged and/or rotating black holes in string theory, the geometry seems to do precisely that -- just before the location of the would-be horizon, the geometry caps off 
(for reviews, see for example~\cite{Bena:2007kg,Gibbons:2013tqa})  
in a jumble of gauge field fluxes and higher-dimensional topology.  In this respect, the geometry seems to be a static version of the Kaluza-Klein `bubble of nothing'%
~\cite{Witten:1981gj}
where the cap of the geometry travels outward along a null surface, but here the null surface is stationary with respect to asymptotic observers rather than expanding out to larger radius.  These horizonless solutions are individual contributors to the ensemble of extremal black holes known as {\it microstate geometries}.

\begin{figure}
\centerline{\includegraphics[width=2in]{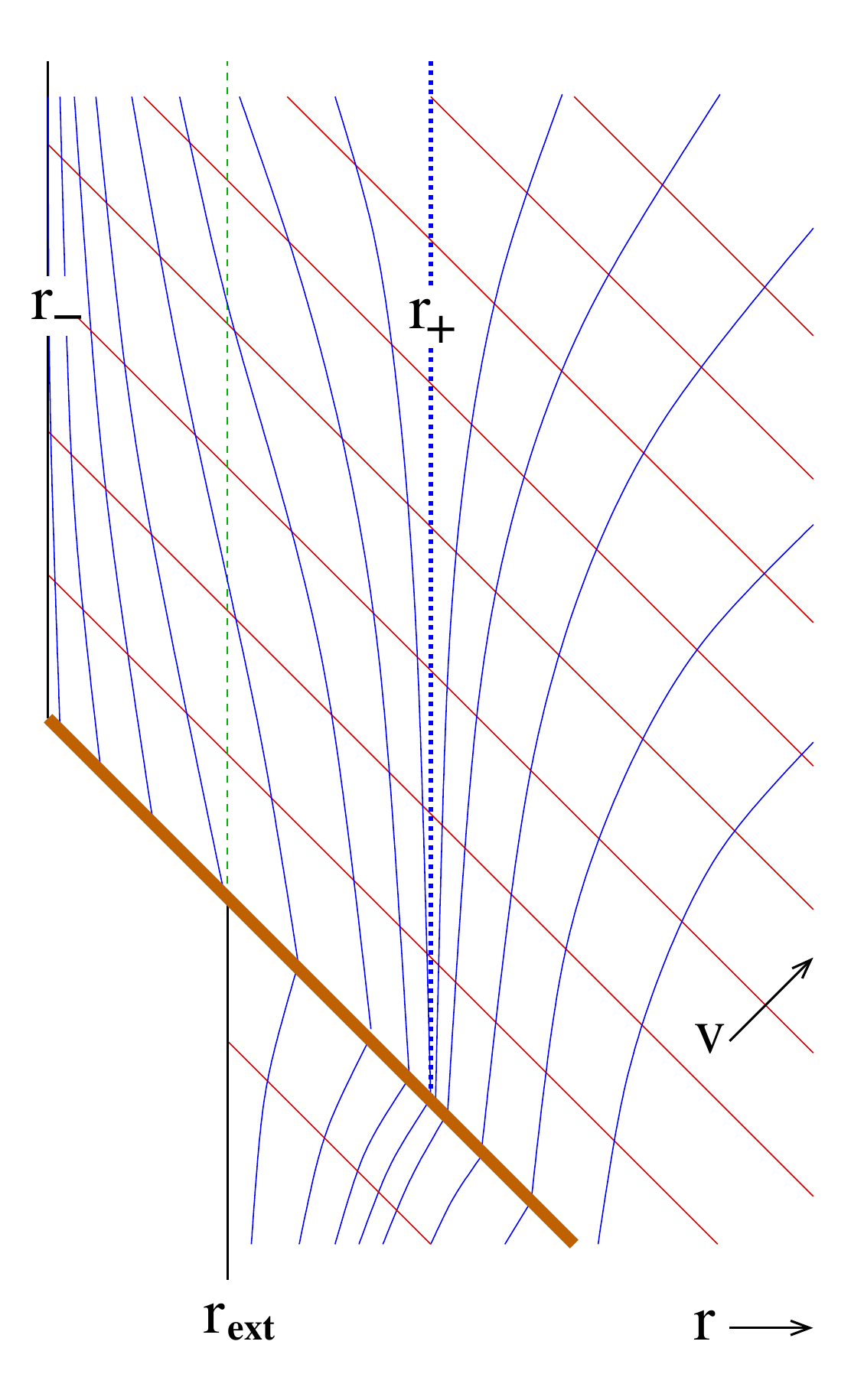}}
\setlength{\unitlength}{0.1\columnwidth}
\caption{\it 
Light cone structure of a shockwave excitation of the extremal black hole geometry. 
}
\label{fig:shockwave}
\end{figure}

In the classical (super)gravity which governs these geometries, the low-energy effective dynamics obeys the strong energy condition; a modest perturbation from extremality results in the formation of a closed trapped surface.  
The cap of a microstate geometry is pushed to smaller radius by such a perturbation, while the outer horizon resides in a locally smooth vacuum geometry and moves out to larger radius.%
\footnote{It seems to be a general feature that the product of the inner and outer horizon areas is a constant independent of the amount of nonextremality or other details of the system%
~\cite{Cvetic:2010mn}.
}
An illustrative example is the infalling shell depicted in figure~\ref{fig:shockwave}.

If horizon thermodynamics is to be believed, there is an entropy $S_-$ associated to the inner horizon of a nonextremal geometry.  Due to the instability of the inner horizon, these degrees of freedom should be associated to, and responsible for resolving, the singularity that appears there according to low-energy effective field theory.  The known extremal microstate geometries seem to have this property, 
and it is tempting to believe that some version of the microstate geometry's cap is operative at the inner horizon of the nonextremal geometry.  
On the other hand, the black hole has a total entropy $S_+$ associated to the area of the outer horizon.  Where is the remaining entropy $\Delta S = S_+-S_-$?  It seems natural to associate this entropy to the inter-horizon region; however that interpretation seems problematic -- no matter having a causal equation of state can remain stationary in the inter-horizon region.


\section{Fractionated Branes}

In string theory, there is substantial evidence that the entropy of black holes is accounted for as the density of states of a new phase of matter whose constituents are fractionated branes.
A standard example is the effective `long string' that accounts for the entropy of black holes in asymptotically $AdS_3\times\bS^3$ spacetime as a bound state of fractionated fivebrane, one-brane, and momentum charges $(n_5,n_1,n_p)$%
~\cite{Strominger:1996sh,Maldacena:1996ds}.
The entropy for $\bS^3$ angular momenta $J_L, J_R$ has the form of a 1+1 dimensional gas
\be
\label{longstring}
S_\pm = 2\pi\sqrt{n_1n_5 (E+n_p)/2-J_L^2} \pm 2\pi\sqrt{n_1n_5(E-n_p)/2-J_R^2} ~.
\ee
A related example is the fractional tension `little string', whose Hagedorn density of states carries the entropy of $n_5$ non-extremal fivebranes%
~\cite{Maldacena:1996ya}
\be
\label{littlestring}
S_\pm = 2\pi\sqrt{n_5 N_L-J_L^2} \pm 2\pi\sqrt{n_5N_R-J_R^2}  ~,
\ee
where $N_L, N_R$ are the left/right excitation levels of the little string. 

In perturbative string theory, there is a correspondence principle crossover%
~\cite{Horowitz:1996nw}
where black holes turn into string states.  In the above systems involving fivebranes, the effective string composed of fractionated branes lies precisely at its correspondence point.  
The entropy~\pref{longstring}
can either be interpreted as the density of states of BTZ black holes in a unitary theory of gravity in a weakly curved $AdS_3\times \bS^3$ spacetime with $\lstr = 4n_1n_5 G_3$, or as the density of states on a long string whose excitations have central charge $c=6$ and a tension reduced by a factor $n_1n_5$.
At weak coupling, this long effective string can be seen quite explicitly in the brane bound state structure; in D1-D5 BPS geometries at strong coupling, it is seen in the (unsmeared) locus of the charge source%
~\cite{Lunin:2002bj}.  
Similarly, the entropy~\pref{littlestring} can be regarded as either that of a linear dilaton black hole in a fivebrane throat with radius $\ell=\sqrt{n_5}\,\lstr$; or as the Hagedorn density of states of the little string whose tension is reduced by a factor of $n_5$.
Despite their disparate appelations, `little strings' and `long strings' are close cousins.

On the stringy side of the correspondence crossover for fundamental strings in asymptotically $AdS_3$ or linear dilaton spacetimes, black holes disappear from the spectrum entirely as one tunes the curvature radius to be smaller than the string length $\lstr$%
~\cite{Giveon:2005mi}.
In this situation, arbitrarily high mass states are fundamental strings rather than black holes.  Horizons or singularities do not form in energetic collisions, rather one makes highly excited strings whose asymptotic density of states has $c_{\rm eff}< 6$; at the correspondence point, $c_{\rm eff}=6$, just like the long/little string.  It is natural to propose that the long/little string that holds the Bekenstein-Hawking entropy doesn't see a black hole as such even if fundamental strings do, because their highly reduced effective tension puts them in the correspondence phase.
The long/little string thus does not respond to the ambient geometry by collapsing into a singularity, any more than a fundamental string at its correspondence point collapses into a black hole; instead, long/little strings see an effectively stringy geometry 
and behave entirely differently by not experiencing a causal structure with a horizon.  The ``horizon'' of the black hole is the extent of the long string state%
~\cite{Mathur:1997wb}.
The ``inter-horizon'' region is then described by a coupled two-phase system -- a Hagedorn gas of the long/little string, interacting with infalling ordinary strings.




\section{Locality, causality, or unitarity -- what to give up?}

\begin{figure}
\centerline{\includegraphics[width=4in]{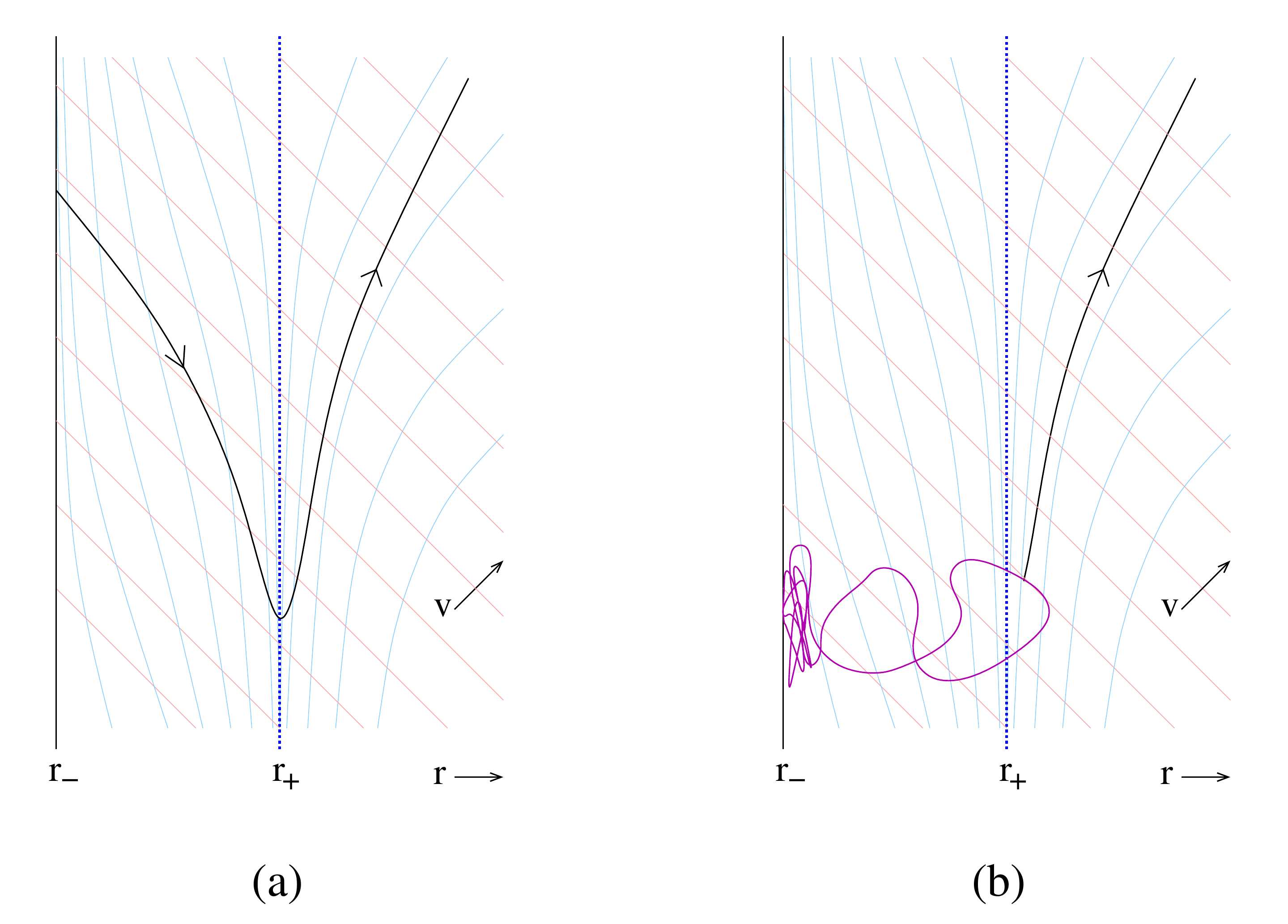}}
\setlength{\unitlength}{0.1\columnwidth}
\caption{\it 
Two descriptions of the Hawking process:  
(a) Pair production in the naive vacuum geometry.  
(b) Radiation from the long/little string.  
}
\label{fig:HawkingRad}
\end{figure}

Hawking's original calculation of black hole radiance%
~\cite{Hawking:1976ra}
revealed the incompatibility of locality, causality, and unitarity properties of effective quantum field theory in the presence of black hole horizons.
Information seems to fall freely into the singularity, where it is spacelike separated from the incoherent pair creation at the outer horizon which generates Hawking radiation, see figure~\ref{fig:HawkingRad}a.  Hawking quanta must radiate that information back to asymptotic observers if unitarity is to be preserved, but that would violate locality and/or causality.
Recently, this paradox has been sharpened via considerations of quantum entanglement%
~\cite{Braunstein:2009v1,Mathur:2009hf,Almheiri:2012rt}. 
If black hole dynamics is unitary, as is suggested by non-perturbative string theory constructions, then Hawking radiation cannot be entangled with partner quanta localized behind the horizon; the black hole interior then cannot be the field theory vacuum inside the horizon, and freely falling observers must encounter violent processes -- a state which has been dubbed a ``firewall".

This conclusion is predicated on an assumption that Hawking quanta are entangled with fluctuations of a field theoretic nature, rather than excitations of the effective long/little string.  Such a string state occupying the black hole interior out to the outer horizon radiates in the manner shown in figure~\ref{fig:HawkingRad}b.  The experience of a freely falling observer is an issue of their interaction with a very low tension object that fills the inter-horizon volume.  Strong interaction with a sufficiently dense fractionated brane soup leads to an evolution akin to meeting a firewall or brick wall -- the long/little string would be a physical realization of the firewall.

At the opposite extreme, sufficiently weak interactions with a very floppy and diffuse effective string leads to a result more akin to a heavy tension D-brane plunging into a Hagedorn gas of fundamental strings, which the heavy brane finds difficult to distinguish from the vacuum over modest time scales.
In this circumstance, infall would be relatively smooth and uneventful until the observer hit the null singularity at the inner horizon.  There, the coupling to long strings grows large; tidal forces rip ordinary matter apart and fractionate it, at which point it joins the long string sector.  Over somewhat longer time scales, the information carried by the infalling object spreads over the long string, migrates to the outer horizon, and is re-radiated as Hawking quanta.  

Either way, the presence of a very light tension fractionated brane state modifies the notion of locality, and the scale of the breakdown of low-energy effective field theory is the horizon scale.


\vskip 2cm




\bibliographystyle{JHEP}
\bibliography{LongStrings}

\end{document}